\documentclass[twocolumn, prl, superscriptaddress,notitlepage]{revtex4-2}
\usepackage{graphicx}
\usepackage{dcolumn}
\usepackage{bm}
\usepackage[usenames,dvipsnames]{color}
\usepackage[most]{tcolorbox}
\usepackage{multirow}
\usepackage{gensymb}
\usepackage[normalem]{ulem}
\usepackage{CJK}
\usepackage{comment}
\usepackage[colorlinks, linkcolor=blue,anchorcolor=blue,citecolor=blue,urlcolor=blue]{hyperref}
\usepackage{amssymb}
\usepackage{pifont}
\usepackage{physics}
\usepackage{natbib}
\usepackage{xcolor}

\usepackage{color,soul}

\begin{document}
\begin{CJK*}{UTF8}{}
\title{Emulated nuclear spin gyroscope with $^{15}$NV centers in diamond
}

\author{Guoqing Wang \CJKfamily{gbsn}(王国庆)}\email[]{gq\_wang@mit.edu}
\thanks{These authors contributed equally.}
\affiliation{
   Department of Nuclear Science and Engineering, Massachusetts Institute of Technology, Cambridge, MA 02139, USA}
\affiliation{
   Research Laboratory of Electronics, Massachusetts Institute of Technology, Cambridge, MA 02139, USA}
\affiliation{Department of Physics, Massachusetts Institute of Technology, Cambridge, MA 02139, USA}

\author{Minh-Thi Nguyen}\email[]{minhthin@mit.edu}
\thanks{These authors contributed equally.}
\affiliation{
   Research Laboratory of Electronics, Massachusetts Institute of Technology, Cambridge, MA 02139, USA}
\affiliation{Department of Physics, Massachusetts Institute of Technology, Cambridge, MA 02139, USA}

\author{Dane W. deQuilettes}
\affiliation{Lincoln Laboratory, Massachusetts Institute of Technology, Lexington, MA 02421, USA}

\author{Eden Price}
\affiliation{Lincoln Laboratory, Massachusetts Institute of Technology, Lexington, MA 02421, USA}

\author{Zhiyao Hu}
\affiliation{
   Research Laboratory of Electronics, Massachusetts Institute of Technology, Cambridge, MA 02139, USA}

\author{Danielle A. Braje}
\affiliation{Lincoln Laboratory, Massachusetts Institute of Technology, Lexington, MA 02421, USA}

\author{Paola Cappellaro}\email[]{pcappell@mit.edu}
\affiliation{
   Department of Nuclear Science and Engineering, Massachusetts Institute of Technology, Cambridge, MA 02139, USA}
\affiliation{
   Research Laboratory of Electronics, Massachusetts Institute of Technology, Cambridge, MA 02139, USA}
\affiliation{Department of Physics, Massachusetts Institute of Technology, Cambridge, MA 02139, USA}

\begin{abstract}
Nuclear spins in solid-state platforms are promising for building rotation sensors due to their long coherence times. Among these platforms, nitrogen-vacancy centers have attracted considerable attention with ambient operating conditions. However, the current performance of NV gyroscopes remains limited by the degraded coherence when operating with large spin ensembles. Protecting the coherence of these systems requires a systematic study of the coherence decay mechanism. Here we present the use of nitrogen-15 nuclear spins of NV centers in building gyroscopes, benefiting from its simpler energy structure and vanishing nuclear quadrupole term compared with nitrogen-14 nuclear spins, though suffering from different challenges in coherence protection. We systematically reveal the coherence decay mechanism of the nuclear spin in different NV electronic spin manifolds and further develop a robust coherence protection protocol based on controlling the NV electronic spin only, achieving a 15-fold dephasing time improvement. With the developed coherence protection, we demonstrate an emulated gyroscope by measuring a designed rotation rate pattern, showing an order-of-magnitude sensitivity improvement. 
\end{abstract}

\maketitle

\end{CJK*}	



\textit{Introduction.---} Sensitive inertial sensors find prevalence in a wide range of applications, from advanced navigation systems~\cite{escobar-alvarez_r-advance_2018} to studies of fundamental physics~\cite{terrano_comagnetometer_2021, smiciklas_new_2011}.  A gyroscope is an instrument that measures rotation relative to an inertial reference frame.  Over the last century, there has been an explosion in innovation of high-performance gyroscopes that take advantage of mechanical sensing \cite{passaro_gyroscope_2017}, the Sagnac effect \cite{arditty_sagnac_1981}, atom interferometry \cite{fang_metrology_2016}, and nuclear spin precession \cite{woodman_nuclear_1987, kornack_nuclear_2005, ajoy_stable_2012}. 
With long coherence times and weak environment coupling, nuclear spins in solid-state platforms have become attractive candidates for gyroscopes~\cite{ajoy_stable_2012,jarmola_demonstration_2021,soshenko_nuclear_2021,wang_characterizing_2023,zhao_inertial_2022,wang_hyperfine-enhanced_2023}. 
The principle of most existing nuclear spin gyroscopes relies on monitoring with Ramsey experiments the  change of the Larmor precession frequency due to the rotation.  To achieve a competitive sensitivity, large ensembles of spins are needed, requiring the use of high spin density samples and large laser power. In turns, this leads to increased magnetic noise due to the spin bath and variations of the intrinsic quadrupole or hyperfine interactions induced, e.g., by temperature and strain~\cite{wang_characterizing_2023}. Thus, protecting the Ramsey coherence of a dense nuclear spin ensemble while maintaining its sensing capabilities is a major task in the field.

Existing solid-state spin-based gyroscopes are mostly based on the spin of the $^{14}$N nucleus ($I=1$) associated with the Nitrogen-Vacancy (NV) center in diamond~\cite{ajoy_stable_2012,jaskula_cross-sensor_2019,jarmola_demonstration_2021,soshenko_nuclear_2021}. Its nuclear quadrupole term sets the spin quantization axis to the defect's orientation along the $z$ direction, allowing the sensing of $z$ rotations. The variations of intrinsic quadrupole and hyperfine interactions typically contribute more than magnetic noise to the spin dephasing, especially in dense spin ensembles~\cite{wang_characterizing_2023}. Thus, to achieve long dephasing times for better gyroscope performance, various methods have been developed to cancel these noise sources. When the nuclear spin probe state is prepared to a single-quantum superposition state $(\ket{0}+\ket{1})/\sqrt{2}$, the recently developed unbalanced echo technique~\cite{wang_characterizing_2023} can refocus the variation of quadrupole term using the hyperfine interaction, requiring only a single quantum $\pi$ pulse on the electronic spin. Alternatively, one could directly prepare the nuclear spin in its double-quantum superposition state $(\ket{+1}+\ket{-1})/\sqrt{2}$, which is immune to both quadrupole and hyperfine variations when NV electronic spin is in $\ket{m_S=0}$ state~\cite{jarmola_demonstration_2021,soshenko_nuclear_2021}.

In comparison to $^{14}$N, the nuclear spin of $^{15}$N is a spin-1/2 without a quadrupole term, and thus in the absence of the hyperfine interaction (electronic spin in its state $|m_S=0\rangle$) its Hamiltonian is independent of the crystal orientation and its dephasing is only limited by the magnetic noise due to the spin bath, 
making it an ideal platform for rotation sensing. Moreover, with a simpler internal structure, the nuclear spin of $^{15}$N is easier to initialize, control, and read out. 
However, the challenges of using $^{15}$N for rotation sensing remain unknown, and an in-depth study of a practical sensing protocol is lacking.

In this work, we explore the use of the $^{15}$N nuclear spin of NV centers in rotation sensing with a diamond sample grown with $^{15}$N isotopic enrichment, by systematically studying the nuclear spin dephasing mechanisms.
First, we reveal that for the NV spin state $\ket{m_S=0}$ the hyperfine interaction enhances the transverse magnetic noise, thus largely degrading the nuclear spin coherence. 
In contrast, the NV states $\ket{m_S=\pm 1}$ suppress this magnetic noise thanks to the large quantization field along $z$ from the hyperfine interaction, whose variations, however, now dominate the dephasing, as  previously seen in $^{14}$N. To circumvent these deleterious effects, we propose to flip the NV spin state between $\ket{m_S=\pm 1}$ during the free evolution of the nuclear spin and demonstrate a 15-fold dephasing time ($T_{2n}^*$) improvement, comparable to the unbalanced echo protection for $^{14}$N nuclear spins~\cite{wang_characterizing_2023}. With the developed protection protocol, we demonstrate an emulated gyroscope by measuring the phase induced by a time-dependent rotation rate and show an order-of-magnitude sensitivity improvement compared to the unprotected case. Our work paves the way to building robust and stable spin-based gyroscopes and long-lived quantum memories using nuclear spin ensembles in solid-state systems.

\begin{figure}[h]
\includegraphics[width=0.5\textwidth]{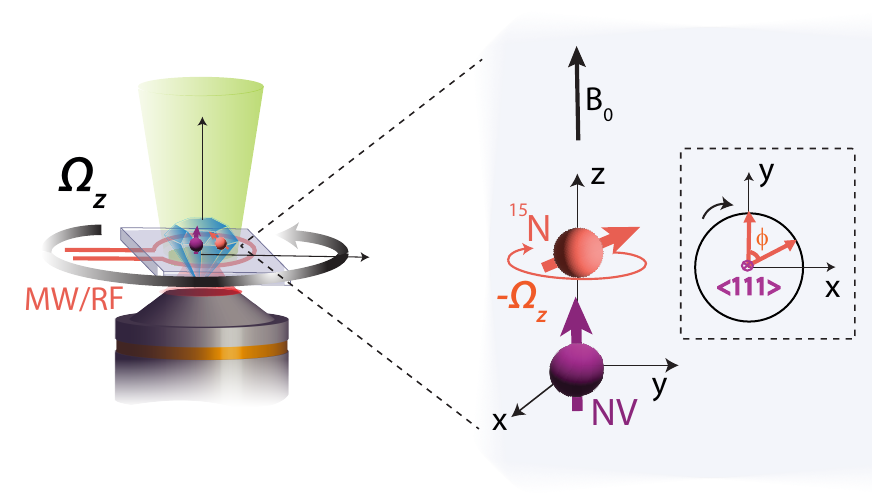}
\caption{\label{fig:1} {\textbf{Schematic of rotation sensor based on $^{15}$NV centers in diamonds.}} Here ``NV" represents the electronic spin of the NV center, while $^{15}$N represents the nuclear spin of the nitrogen as part of the NV. Note that in this work we do not rotate the diamond but instead emulate the gyroscope by varying the phase of the last $\pi/2$ pulse in the Ramsey sequence.}
\end{figure}

\begin{figure*}[htbp]
\includegraphics[width=\textwidth]{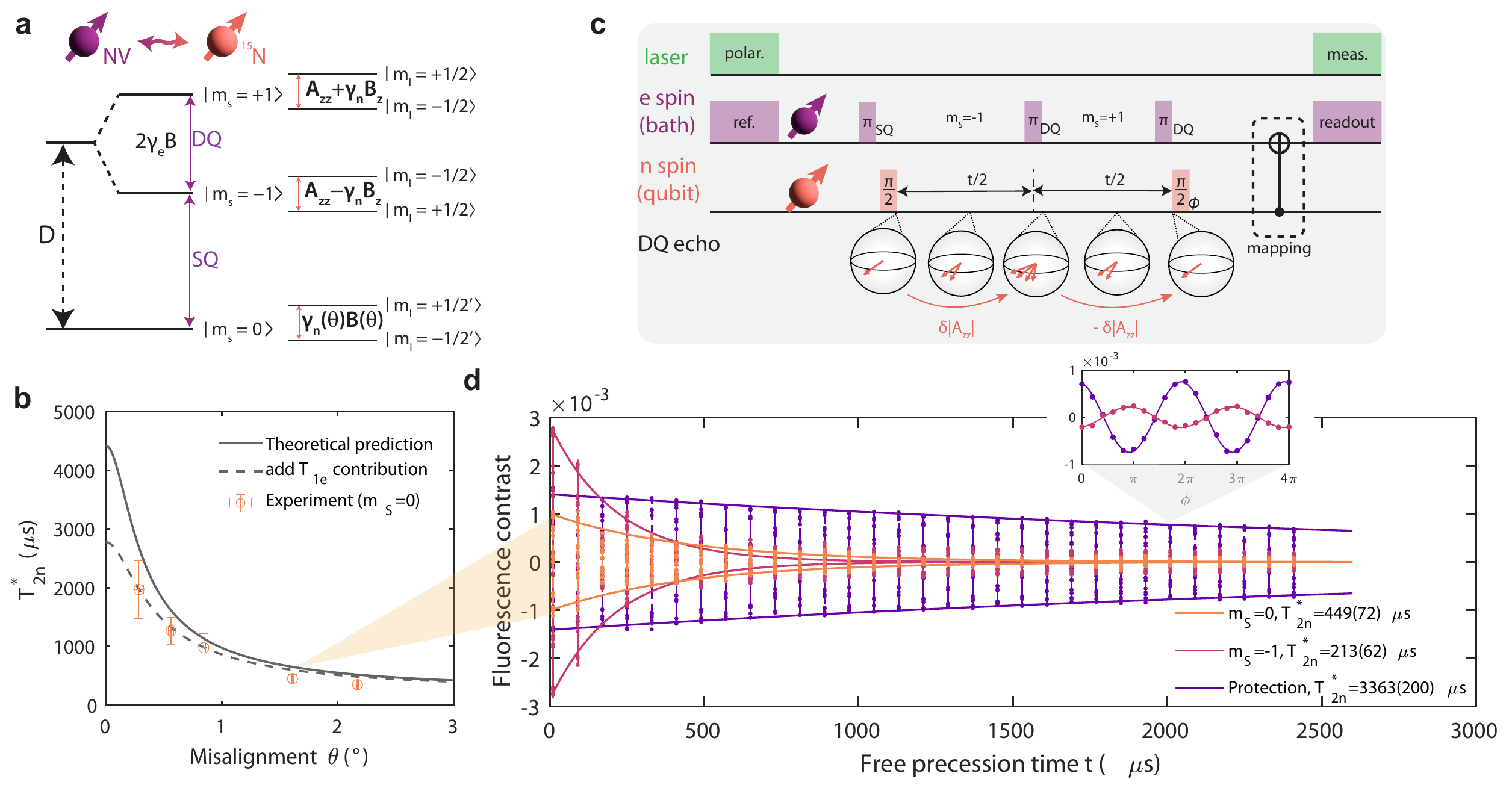}
\caption{\label{fig:2} \textbf{Dephasing mechanism and coherence protection.} (a) Energy structure of the $^{15}$NV center. (b) Dephasing measurement under different misalignment angles and theoretical predictions. The theoretical prediction is based on directly diagonalizing the full Hamiltonian of the NV-nuclear spin system assuming a 239~Gauss magnetic field is misaligned from the NV axis by an angle $\theta$, with $1/T_{2n}^*=1/T_{2n,B}^*+1/(1.5T_{1e})$ shown by the dashed gray line and $1/T_{2n,B}^*$ the prediction based on magnetic noise only indicated by the solid gray line. (c) Coherence protection sequence. (d) Nuclear spin dephasing time $T_{2n}^*$ measurement with and without the protection sequence under a misalignment angle of about $\theta\approx 1.6\degree$ indicated by the orange shadow pointing to (b). Phase sweep of the last $\pi/2$ pulse at each evolution time is performed to characterize the coherence. The dephasing time is fitted from an exponential decay shape $S(t)=ce^{-t/T_{2n}^*}$. 
}
\end{figure*}

\textit{Spin dephasing mechanism.---} The system and principle of the solid-state spin-based rotation sensor are shown in Fig.~\ref{fig:1}. When the sensor (including the host material of the sensor spin and microwave/radiofrequency delivery structures) rotates about the $z$ axis with a rate $\Omega_z$, the inertial nuclear spin-1/2 prepared in a superposition state $(\ket{+1/2}+\ket{-1/2})/\sqrt{2}$ counter-rotates with respect to the host material and accumulates a phase $\phi=\Omega_z t$. The rotation rate can then be extracted by measuring the phase $\phi$ through, e.g., Ramsey interferometry.

The experimental platform we study is based on NV centers in diamonds.
The NV center is an atom-like spin defect consisting of an electronic spin-1 and a nitrogen nuclear spin-1/2. The electronic spin state $\ket{m_S=0}$ of the NV center can be prepared and read out by illuminating the diamond with a green laser, while the nuclear spin state can be prepared by transferring polarization from the electronic spin and read out by mapping the spin state to the electronic spin~\cite{doherty_nitrogen-vacancy_2013}.  When the $z$ axis of the reference frame is chosen to be along the NV orientation, the ground state Hamiltonian of the NV center can be written as
\begin{equation}
    H = DS^2_z + \gamma_e \Vec{B} \cdot \Vec{S} + \gamma_n \Vec{B} \cdot \Vec{I} + \Vec{S} \cdot \bold{A} \cdot \Vec{I}
\end{equation}
where $D/(2\pi)=2.87$~GHz is the zero-field splitting (ZFS) of NV electronic spin, $\gamma_e/(2\pi)=2.8024$~MHz/G and $\gamma_n/(2\pi)=0.4316$~kHz/G are the gyromagnetic ratios of the electronic and nuclear spin, $\bold{A}$ is the hyperfine tensor including only diagonal terms $A_{zz}/(2\pi)=3.03$~MHz and $A_{\perp}/(2\pi)=3.65$~MHz. When the external magnetic field $\Vec{B}$ is small ($\gamma_eB\ll D$), the electronic spin is quantized along the NV axis by the large ZFS term, thus is tied to the diamond crystal. In contrast, the nuclear spin has an effective Hamiltonian~\cite{chen_measurement_2015,sangtawesin_hyperfine-enhanced_2016,oon_ramsey_2022-1}
\begin{equation}
\label{eq:H_I}
    H_I=A_{zz}m_s{I}_z +\gamma_n\left[B_z{I}_z +\alpha_{m_s}  B_x {I}_x\right],
\end{equation}
where $\alpha_{m_s}$ is an enhancement factor of the transverse Zeeman coupling induced by the transverse hyperfine interaction~\cite{chen_measurement_2015,sangtawesin_hyperfine-enhanced_2016}. For small external magnetic fields  ($\gamma_eB\ll D$), $\alpha_{m_s}$ can be approximated to a constant $\alpha=(1-2\kappa+3\kappa m_s^2)$, with $\kappa\approx\gamma_eA_\perp/(\gamma_n D)\approx 8.26$~\cite{SOM,oon_ramsey_2022-1}.

When the NV spin state is $|m_S=0\rangle$, the enhancement of the transverse Zeeman coupling increases the transverse magnetic noise by about an order of magnitude. Such an enhancement leads to faster dephasing rates when the field alignment to the NV axis is imperfect or the field direction is inhomogeneous. 
Quantitatively, for a field misalignment $\theta$ from the NV z-axis, the nuclear spin transition frequency is $\omega_n(\theta)=\gamma_nB\sqrt{\cos^2\theta+\alpha_0^2\sin^2\theta}$. Assuming the longitudinal and transverse magnetic noise are independent and both have a mean strength $\Delta_B=\sqrt{\langle\delta B_{z,x}^2\rangle}$, the propagated variance of the nuclear spin transition frequency is then $\langle\delta\omega_n^2\rangle=\gamma_n^2\Delta_B^2f(\theta)^2$ with a noise enhancement factor $f(\theta)=\sqrt{(1+\alpha_0^4\tan^2\theta)/(1+\alpha_0^2\tan^2\theta)}$. The magnetic noise-limited dephasing time is thus $T_{2n}^*=1/(\gamma_n\Delta_B f(\theta))$. 
Even for a small misalignment the enhancement can be quite large (e.g., when $\theta=1\degree$, $f(\theta)=4.17$). We note that this sensitive effect is due to the nearly 15-fold transverse magnetic field enhancement which amplifies both the transverse magnetic noise and the quantization axis tilt due to misalignment.

The strength of magnetic noise can be characterized by the NV dephasing time $T_{2e}^*=1/(\gamma_e\Delta_B)=0.69~\mu$s for our sample~\cite{SOM}, from which we can predict a magnetic noise-limited dephasing time $T_{2n}^*\approx(1/f(\theta))4.48$~ms. 
To validate our theoretical prediction, we measure $T_{2n}^*$ as a function of field misalignment $\theta$, which is evaluated by the nuclear spin transition frequency $\omega_n(\theta)$~\cite{SOM}. The experimental results shown in Fig.~\ref{fig:2}b are consistent with the trend of the theoretical prediction, while the measured dephasing times are slightly shorter due to the existence of other dephasing mechanisms. One potential contribution is the NV spin relaxation time $T_{1e}\approx 5$~ms that can lead to an additional dephasing rate $1/(1.5T_{1e})$~\cite{chen_protecting_2018}. When taking this into account (dashed line in Fig.~\ref{fig:2}b) the theoretical prediction better matches the experimental result.

\textit{Coherence protection.---} To eliminate the enhanced transverse magnetic noise contribution, one can work with NV spin states $\ket{m_S=\pm1}$ which apply a large longitudinal quantization field $A_{zz}$ (equivalent to 0.7~T) to the nuclear spin. In these conditions, the nuclear spin energy splitting is no longer affected by the relatively small transverse magnetic field and its noise $\omega_n(m_S=\pm 1)\approx\gamma_n B \pm A_{zz}$. Ideally, the dephasing time dominated by the longitudinal magnetic noise should reach a few milliseconds as predicted by the previous analysis. However, the observed dephasing time is only about 200~$\mu$s (Fig.~\ref{fig:2}d), one order of magnitude smaller and even shorter than for $m_S=0$. The observation is consistent with our previous study~\cite{wang_characterizing_2023}, which shows that the  dephasing is now dominated by the variations of the hyperfine interaction due to inhomogeneity or fluctuation of temperature, strain, electric field, etc. The identification of noise sources and their effects on $A_{zz}$ is interesting and nontrivial, but requires more independent characterizations combined with first-principles calculations~\cite{wang_characterizing_2023,tang_first-principles_2023}. In this work, we control the laser heating to vary the sample temperature and characterize temperature shifts of the hyperfine interaction near room temperature and obtain $dA_{zz}/dT=-272(8)$~Hz/K, consistent with the recent experimental studies~\cite{lourette_temperature_2022}. Our temperature-dependence measurement details are included in the supplemental materials~\cite{SOM}.

To refocus the variation of hyperfine interaction while still maintaining nuclear spin's free evolution, we can apply a double-quantum (DQ) $\pi$ pulse to flip the electronic spin state from $\ket{m_S=-1}$ to $\ket{m_S=+1}$. The protection sequence is shown in Fig.~\ref{fig:2}c. The effective hyperfine field on the nuclear spin, as well as its variations $\delta A_{zz}$, are canceled, leaving the nuclear spin to evolve under the longitudinal magnetic field component. As shown in Fig.~\ref{fig:2}d, the nuclear spin dephasing time under the DQ protection sequence is extended to about 3.4~ms, more than a factor of 15 longer than the unprotected case with the NV in its state $\ket{m_S=-1}$. The complete dephasing time measurements with and without the protection under different misalignment angles are included in the supplemental materials~\cite{SOM}, where more than 3~ms dephasing times are obtained when applying the coherence protection protocol in all misalignment angles we measured.

\begin{figure}[h]
\includegraphics[width=0.5\textwidth]{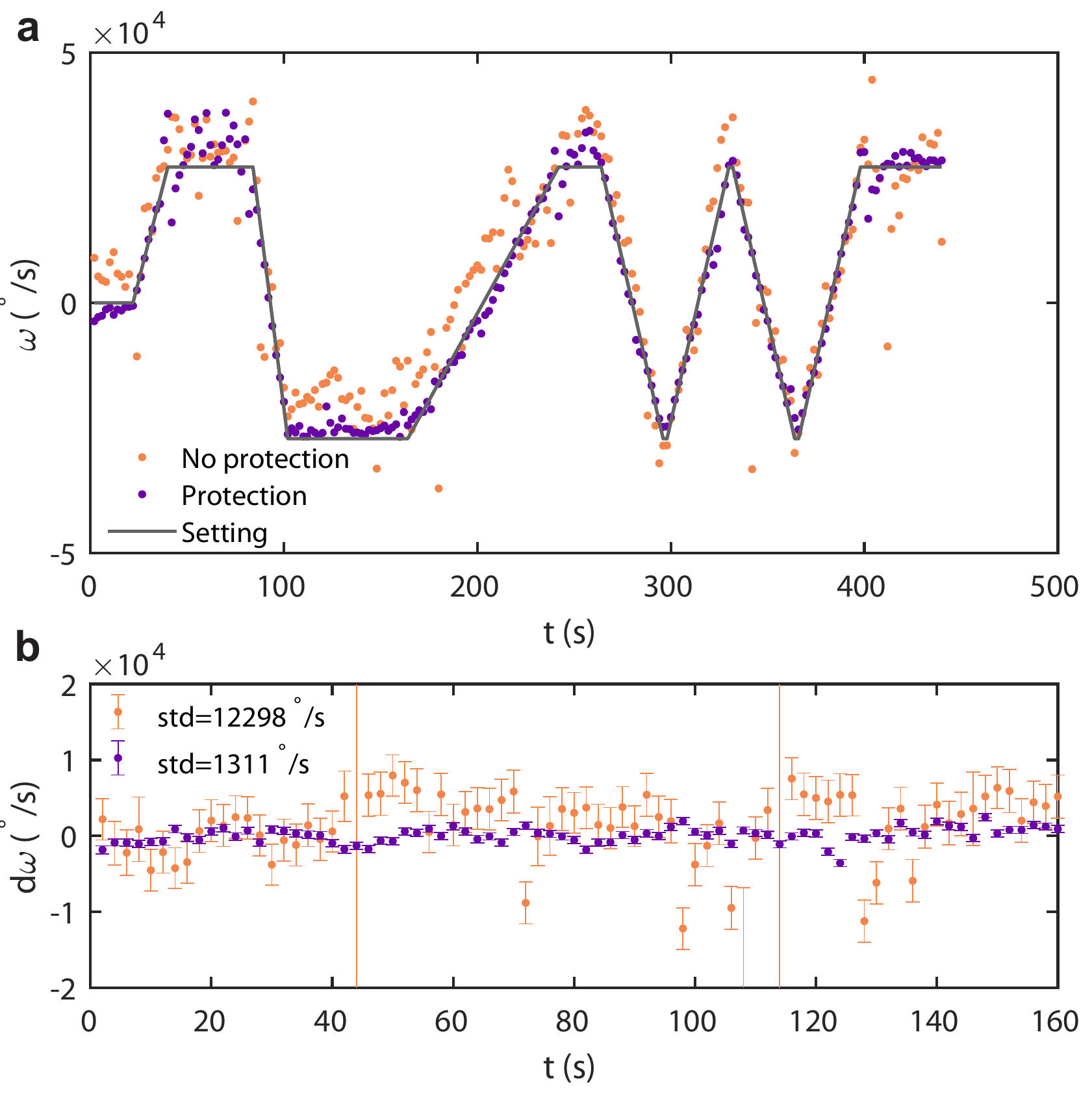}
\caption{\label{fig:3} \textbf{Demonstration of an emulated gyroscope.} (a) Rotation rate $\omega$ setting and measurement with and without protection sequence. Evolution times for the protection and no protection ($m_S=-1$) cases are chosen to be $2.2,0.2$~ms, respectively, while the dead time of both sequences is about $1.8,1.25$~ms. Thus the repetition numbers are chosen to be 400 and 690 such that the measurement for a single data point takes 2~s. (b) Rotation rate uncertainty at zero-bias point $\omega=0$ obtained from a separate measurement. The standard deviation of the two methods shows the uncertainty of the gyroscope measurement. }
\end{figure}

\textit{Emulated gyroscope.---} The protected nuclear spin sensors can now be used to detect rotations along the NV z-axis, in the presence of a(n average) magnetic field along $z$ as well.
To demonstrate a rotation sensing and benchmark the performance of the protection protocol based on DQ control, we modulate the phase $\varphi$ of the last $\pi/2$ pulse in the nuclear spin Ramsey sequence to emulate a gyroscope. The rotation rate $\Omega_z$ to be measured can be extracted from the accumulated phase  $\phi=\Omega_z t$. We calibrate a constant phase offset to cancel out the phase induced by the longitudinal magnetic field and imperfect DQ $\pi$ pulse locations. We further apply a 2-Ramsey protocol by alternating the phase of the last $\pi/2$ pulse between $\varphi$ and $\varphi+\pi$, such that the measured signal has a form $S=c_0\pm c\sin(\Omega_z t)$ with $c_0$ the offset and $c$ the contrast. By subtracting the two signals, any common-mode drifts in the offset $c_0$ are canceled. We set a phase pattern to mimic a time-dependent rotation which is measured with and without coherence protection, as shown in Fig.~\ref{fig:3}a. The repetitions of experimental sequences take 2~s for each data point in the plot. The gyroscope measurement under DQ protection is less noisy than the one without protection. We then benchmark the sensitivity of both cases by measuring the standard deviation of the reconstructed rotation rate at the zero-bias point $\Omega_z=0$. The experiment under the protection shows a standard deviation of 1311~\degree/s, one order of magnitude better than the unprotected case 12298~\degree/s. Correspondingly, the achieved sensitivity under the protection is then 927~\degree/s Hz$^{-1/2}$. 

The sensitivity improvement under the DQ control comes from not only the coherence protection but also the improved sensor stability against drifts in experimental conditions (e.g., temperature, magnetic field, etc.). Random temperature drifts can introduce additional phases in the Ramsey experiments, leading to a systematic bias of the rotation sensor as shown in Fig.~\ref{fig:3}b. 
For the NV $\ket{m_S=0}$ nuclear spin manifold, drifts in magnetic alignment not only lead to degraded coherence but also  to lower signal contrast due to the quantization axis change and worse initialization/readout. 
The sensing protocol based on DQ protection is immune to these experimental condition drifts, and by using RF sequences to polarize the nuclear spin, the signal contrast is comparable with experiments on the NV electronic spin (see supplement \cite{SOM}).

\textit{Conclusion and discussions.---} In this work we considered the $^{15}$N spins associated with NV centers in diamond as potential rotation sensors. 
With a systematic study of the nuclear spin dephasing mechanism in different electronic spin manifolds, we develop and demonstrate the optimal Ramsey-type gyroscope protocol for these nuclear spin-1/2. 
Our findings indicate that in the electronic spin $m_S=0$ manifold, the nuclear spin within is more susceptible to  magnetic noise, which is enhanced by the transverse hyperfine interaction, even under minimal misalignment. Consequently, leaving the NV in the $m_S=0$ state is not a suitable choice to exploit nuclear spins as rotation sensors, particularly when employing dense spin ensembles to improve sensitivity.

A better strategy is to work in the electronic spin states $m_S=\pm1$ and apply a DQ spin flip to decouple the now dominant noise arising from the hyperfine interaction. Then, the dephasing time of the $^{15}$N nuclear spin ensemble can be extended to about 3.4(0.2)~ms, comparable with the $^{14}$N nuclear spin ensemble where a dephasing time of about 3.5(0.4)~ms is achieved under the protection of unbalanced echo in our previous work~\cite{wang_characterizing_2023}, or with double-quantum Ramsey gyroscope protocol demonstrated in Refs.~\cite{jarmola_demonstration_2021,soshenko_nuclear_2021}. Despite the lack of quadrupolar interaction, which is a source of noise for $^{14}$N, the $^{15}$N spin-1/2 coherence time is thus not longer, as one might have naively assumed.
Still, using a spin-1/2 might introduce other practical advantages. In comparison to $^{14}$N, $^{15}$N only has two nuclear spin energy levels and a larger hyperfine interaction strength. Thus, the control and polarization of the nuclear spin are less complicated. In addition, the unbalanced echo protocol developed for $^{14}$N cancels only one major variation source such as temperature or strain given that the unbalanced echo $\pi$ pulse location $\tau/t$ can only be set to a ratio between the variations of quadrupole term $Q$ and hyperfine interaction $A_{zz}$, which is different for different noise sources. Using $^{15}$N instead allows the complete cancellation of the variation in $A_{zz}$, thus the coherence is only limited by the noise of the longitudinal magnetic field.

Our work not only paves the way for building gyroscopes using large ensembles of NV nuclear spins but also helps in developing quantum memories 
useful for quantum error correction, quantum networks, and quantum sensing, which usually require a longer coherence time. The use of enhanced transverse Zeeman coupling allows the characterization of transverse magnetic noise due to the spin bath and can be further exploited in magnetic sensing~\cite{liu_nanoscale_2019}.

\acknowledgements
This work was supported in part by DARPA DRINQS program (Cooperative Agreement No. D18AC00024). G.W. thanks MathWorks for their support in the form of a Graduate Student Fellowship. The opinions and views expressed in this publication are from the authors and
not necessarily from MathWorks.

\bibliography{main_text}

\begin{thebibliography}{25}%
\makeatletter
\providecommand \@ifxundefined [1]{%
 \@ifx{#1\undefined}
}%
\providecommand \@ifnum [1]{%
 \ifnum #1\expandafter \@firstoftwo
 \else \expandafter \@secondoftwo
 \fi
}%
\providecommand \@ifx [1]{%
 \ifx #1\expandafter \@firstoftwo
 \else \expandafter \@secondoftwo
 \fi
}%
\providecommand \natexlab [1]{#1}%
\providecommand \enquote  [1]{``#1''}%
\providecommand \bibnamefont  [1]{#1}%
\providecommand \bibfnamefont [1]{#1}%
\providecommand \citenamefont [1]{#1}%
\providecommand \href@noop [0]{\@secondoftwo}%
\providecommand \href [0]{\begingroup \@sanitize@url \@href}%
\providecommand \@href[1]{\@@startlink{#1}\@@href}%
\providecommand \@@href[1]{\endgroup#1\@@endlink}%
\providecommand \@sanitize@url [0]{\catcode `\\12\catcode `\$12\catcode
  `\&12\catcode `\#12\catcode `\^12\catcode `\_12\catcode `\%12\relax}%
\providecommand \@@startlink[1]{}%
\providecommand \@@endlink[0]{}%
\providecommand \url  [0]{\begingroup\@sanitize@url \@url }%
\providecommand \@url [1]{\endgroup\@href {#1}{\urlprefix }}%
\providecommand \urlprefix  [0]{URL }%
\providecommand \Eprint [0]{\href }%
\providecommand \doibase [0]{https://doi.org/}%
\providecommand \selectlanguage [0]{\@gobble}%
\providecommand \bibinfo  [0]{\@secondoftwo}%
\providecommand \bibfield  [0]{\@secondoftwo}%
\providecommand \translation [1]{[#1]}%
\providecommand \BibitemOpen [0]{}%
\providecommand \bibitemStop [0]{}%
\providecommand \bibitemNoStop [0]{.\EOS\space}%
\providecommand \EOS [0]{\spacefactor3000\relax}%
\providecommand \BibitemShut  [1]{\csname bibitem#1\endcsname}%
\let\auto@bib@innerbib\@empty
\bibitem [{\citenamefont {{Escobar-Alvarez}}\ \emph {et~al.}(2018)\citenamefont
  {{Escobar-Alvarez}}, \citenamefont {Johnson}, \citenamefont {Hebble},
  \citenamefont {Klingebiel}, \citenamefont {Quintero}, \citenamefont
  {Regenstein},\ and\ \citenamefont
  {Browning}}]{escobar-alvarez_r-advance_2018}%
  \BibitemOpen
  \bibfield  {author} {\bibinfo {author} {\bibfnamefont {H.~D.}\ \bibnamefont
  {{Escobar-Alvarez}}}, \bibinfo {author} {\bibfnamefont {N.}~\bibnamefont
  {Johnson}}, \bibinfo {author} {\bibfnamefont {T.}~\bibnamefont {Hebble}},
  \bibinfo {author} {\bibfnamefont {K.}~\bibnamefont {Klingebiel}}, \bibinfo
  {author} {\bibfnamefont {S.~A.~P.}\ \bibnamefont {Quintero}}, \bibinfo
  {author} {\bibfnamefont {J.}~\bibnamefont {Regenstein}},\ and\ \bibinfo
  {author} {\bibfnamefont {N.~A.}\ \bibnamefont {Browning}},\ }\bibfield
  {title} {\bibinfo {title} {R-{{ADVANCE}}: {{Rapid Adaptive Prediction}} for
  {{Vision-based Autonomous Navigation}}, {{Control}}, and {{Evasion}}},\
  }\href {https://doi.org/10.1002/rob.21744} {\bibfield  {journal} {\bibinfo
  {journal} {J. Field Robot.}\ }\textbf {\bibinfo {volume} {35}},\ \bibinfo
  {pages} {91} (\bibinfo {year} {2018})}\BibitemShut {NoStop}%
\bibitem [{\citenamefont {Terrano}\ and\ \citenamefont
  {Romalis}(2021)}]{terrano_comagnetometer_2021}%
  \BibitemOpen
  \bibfield  {author} {\bibinfo {author} {\bibfnamefont {W.~A.}\ \bibnamefont
  {Terrano}}\ and\ \bibinfo {author} {\bibfnamefont {M.~V.}\ \bibnamefont
  {Romalis}},\ }\bibfield  {title} {\bibinfo {title} {Comagnetometer probes of
  dark matter and new physics},\ }\href
  {https://doi.org/10.1088/2058-9565/ac1ae0} {\bibfield  {journal} {\bibinfo
  {journal} {Quantum Sci. Technol.}\ }\textbf {\bibinfo {volume} {7}},\
  \bibinfo {pages} {014001} (\bibinfo {year} {2021})}\BibitemShut {NoStop}%
\bibitem [{\citenamefont {Smiciklas}\ \emph {et~al.}(2011)\citenamefont
  {Smiciklas}, \citenamefont {Brown}, \citenamefont {Cheuk}, \citenamefont
  {Smullin},\ and\ \citenamefont {Romalis}}]{smiciklas_new_2011}%
  \BibitemOpen
  \bibfield  {author} {\bibinfo {author} {\bibfnamefont {M.}~\bibnamefont
  {Smiciklas}}, \bibinfo {author} {\bibfnamefont {J.~M.}\ \bibnamefont
  {Brown}}, \bibinfo {author} {\bibfnamefont {L.~W.}\ \bibnamefont {Cheuk}},
  \bibinfo {author} {\bibfnamefont {S.~J.}\ \bibnamefont {Smullin}},\ and\
  \bibinfo {author} {\bibfnamefont {M.~V.}\ \bibnamefont {Romalis}},\
  }\bibfield  {title} {\bibinfo {title} {New {{Test}} of {{Local Lorentz
  Invariance Using}} a {{Ne}} 21 - {{Rb}} - {{K Comagnetometer}}},\ }\href
  {https://doi.org/10.1103/PhysRevLett.107.171604} {\bibfield  {journal}
  {\bibinfo  {journal} {Phys. Rev. Lett.}\ }\textbf {\bibinfo {volume} {107}},\
  \bibinfo {pages} {171604} (\bibinfo {year} {2011})}\BibitemShut {NoStop}%
\bibitem [{\citenamefont {Passaro}\ \emph {et~al.}(2017)\citenamefont
  {Passaro}, \citenamefont {Cuccovillo}, \citenamefont {Vaiani}, \citenamefont
  {De~Carlo},\ and\ \citenamefont {Campanella}}]{passaro_gyroscope_2017}%
  \BibitemOpen
  \bibfield  {author} {\bibinfo {author} {\bibfnamefont {V.~M.~N.}\
  \bibnamefont {Passaro}}, \bibinfo {author} {\bibfnamefont {A.}~\bibnamefont
  {Cuccovillo}}, \bibinfo {author} {\bibfnamefont {L.}~\bibnamefont {Vaiani}},
  \bibinfo {author} {\bibfnamefont {M.}~\bibnamefont {De~Carlo}},\ and\
  \bibinfo {author} {\bibfnamefont {C.~E.}\ \bibnamefont {Campanella}},\
  }\bibfield  {title} {\bibinfo {title} {Gyroscope {{Technology}} and
  {{Applications}}: {{A Review}} in the {{Industrial Perspective}}},\ }\href
  {https://doi.org/10.3390/s17102284} {\bibfield  {journal} {\bibinfo
  {journal} {Sensors}\ }\textbf {\bibinfo {volume} {17}},\ \bibinfo {pages}
  {2284} (\bibinfo {year} {2017})}\BibitemShut {NoStop}%
\bibitem [{\citenamefont {Arditty}\ and\ \citenamefont
  {Lef{\`e}vre}(1981)}]{arditty_sagnac_1981}%
  \BibitemOpen
  \bibfield  {author} {\bibinfo {author} {\bibfnamefont {H.~J.}\ \bibnamefont
  {Arditty}}\ and\ \bibinfo {author} {\bibfnamefont {H.~C.}\ \bibnamefont
  {Lef{\`e}vre}},\ }\bibfield  {title} {\bibinfo {title} {Sagnac effect in
  fiber gyroscopes},\ }\href {https://doi.org/10.1364/OL.6.000401} {\bibfield
  {journal} {\bibinfo  {journal} {Opt. Lett.}\ }\textbf {\bibinfo {volume}
  {6}},\ \bibinfo {pages} {401} (\bibinfo {year} {1981})}\BibitemShut {NoStop}%
\bibitem [{\citenamefont {Fang}\ \emph {et~al.}(2016)\citenamefont {Fang},
  \citenamefont {Dutta}, \citenamefont {Gillot}, \citenamefont {Savoie},
  \citenamefont {Lautier}, \citenamefont {Cheng}, \citenamefont {Alzar},
  \citenamefont {Geiger}, \citenamefont {Merlet}, \citenamefont {Santos},\ and\
  \citenamefont {Landragin}}]{fang_metrology_2016}%
  \BibitemOpen
  \bibfield  {author} {\bibinfo {author} {\bibfnamefont {B.}~\bibnamefont
  {Fang}}, \bibinfo {author} {\bibfnamefont {I.}~\bibnamefont {Dutta}},
  \bibinfo {author} {\bibfnamefont {P.}~\bibnamefont {Gillot}}, \bibinfo
  {author} {\bibfnamefont {D.}~\bibnamefont {Savoie}}, \bibinfo {author}
  {\bibfnamefont {J.}~\bibnamefont {Lautier}}, \bibinfo {author} {\bibfnamefont
  {B.}~\bibnamefont {Cheng}}, \bibinfo {author} {\bibfnamefont {C.~L.~G.}\
  \bibnamefont {Alzar}}, \bibinfo {author} {\bibfnamefont {R.}~\bibnamefont
  {Geiger}}, \bibinfo {author} {\bibfnamefont {S.}~\bibnamefont {Merlet}},
  \bibinfo {author} {\bibfnamefont {F.~P.~D.}\ \bibnamefont {Santos}},\ and\
  \bibinfo {author} {\bibfnamefont {A.}~\bibnamefont {Landragin}},\ }\bibfield
  {title} {\bibinfo {title} {Metrology with {{Atom Interferometry}}: {{Inertial
  Sensors}} from {{Laboratory}} to {{Field Applications}}},\ }\href
  {https://doi.org/10.1088/1742-6596/723/1/012049} {\bibfield  {journal}
  {\bibinfo  {journal} {J. Phys. Conf. Ser.}\ }\textbf {\bibinfo {volume}
  {723}},\ \bibinfo {pages} {012049} (\bibinfo {year} {2016})}\BibitemShut
  {NoStop}%
\bibitem [{\citenamefont {Woodman}\ \emph {et~al.}(1987)\citenamefont
  {Woodman}, \citenamefont {Franks},\ and\ \citenamefont
  {Richards}}]{woodman_nuclear_1987}%
  \BibitemOpen
  \bibfield  {author} {\bibinfo {author} {\bibfnamefont {K.~F.}\ \bibnamefont
  {Woodman}}, \bibinfo {author} {\bibfnamefont {P.~W.}\ \bibnamefont
  {Franks}},\ and\ \bibinfo {author} {\bibfnamefont {M.~D.}\ \bibnamefont
  {Richards}},\ }\bibfield  {title} {\bibinfo {title} {The {{Nuclear Magnetic
  Resonance Gyroscope}}: A {{Review}}},\ }\href
  {https://doi.org/10.1017/S037346330000062X} {\bibfield  {journal} {\bibinfo
  {journal} {J. Navig.}\ }\textbf {\bibinfo {volume} {40}},\ \bibinfo {pages}
  {366} (\bibinfo {year} {1987})}\BibitemShut {NoStop}%
\bibitem [{\citenamefont {Kornack}\ \emph {et~al.}(2005)\citenamefont
  {Kornack}, \citenamefont {Ghosh},\ and\ \citenamefont
  {Romalis}}]{kornack_nuclear_2005}%
  \BibitemOpen
  \bibfield  {author} {\bibinfo {author} {\bibfnamefont {T.~W.}\ \bibnamefont
  {Kornack}}, \bibinfo {author} {\bibfnamefont {R.~K.}\ \bibnamefont {Ghosh}},\
  and\ \bibinfo {author} {\bibfnamefont {M.~V.}\ \bibnamefont {Romalis}},\
  }\bibfield  {title} {\bibinfo {title} {Nuclear {{Spin Gyroscope Based}} on an
  {{Atomic Comagnetometer}}},\ }\href
  {https://doi.org/10.1103/PhysRevLett.95.230801} {\bibfield  {journal}
  {\bibinfo  {journal} {Phys. Rev. Lett.}\ }\textbf {\bibinfo {volume} {95}},\
  \bibinfo {pages} {230801} (\bibinfo {year} {2005})}\BibitemShut {NoStop}%
\bibitem [{\citenamefont {Ajoy}\ and\ \citenamefont
  {Cappellaro}(2012)}]{ajoy_stable_2012}%
  \BibitemOpen
  \bibfield  {author} {\bibinfo {author} {\bibfnamefont {A.}~\bibnamefont
  {Ajoy}}\ and\ \bibinfo {author} {\bibfnamefont {P.}~\bibnamefont
  {Cappellaro}},\ }\bibfield  {title} {\bibinfo {title} {Stable three-axis
  nuclear-spin gyroscope in diamond},\ }\href
  {https://doi.org/10.1103/PhysRevA.86.062104} {\bibfield  {journal} {\bibinfo
  {journal} {Phy. Rev. A}\ }\textbf {\bibinfo {volume} {86}},\ \bibinfo {pages}
  {062104} (\bibinfo {year} {2012})}\BibitemShut {NoStop}%
\bibitem [{\citenamefont {Jarmola}\ \emph {et~al.}(2021)\citenamefont
  {Jarmola}, \citenamefont {Lourette}, \citenamefont {Acosta}, \citenamefont
  {Birdwell}, \citenamefont {Bl{\"u}mler}, \citenamefont {Budker},
  \citenamefont {Ivanov},\ and\ \citenamefont
  {Malinovsky}}]{jarmola_demonstration_2021}%
  \BibitemOpen
  \bibfield  {author} {\bibinfo {author} {\bibfnamefont {A.}~\bibnamefont
  {Jarmola}}, \bibinfo {author} {\bibfnamefont {S.}~\bibnamefont {Lourette}},
  \bibinfo {author} {\bibfnamefont {V.~M.}\ \bibnamefont {Acosta}}, \bibinfo
  {author} {\bibfnamefont {A.~G.}\ \bibnamefont {Birdwell}}, \bibinfo {author}
  {\bibfnamefont {P.}~\bibnamefont {Bl{\"u}mler}}, \bibinfo {author}
  {\bibfnamefont {D.}~\bibnamefont {Budker}}, \bibinfo {author} {\bibfnamefont
  {T.}~\bibnamefont {Ivanov}},\ and\ \bibinfo {author} {\bibfnamefont {V.~S.}\
  \bibnamefont {Malinovsky}},\ }\bibfield  {title} {\bibinfo {title}
  {Demonstration of diamond nuclear spin gyroscope},\ }\href
  {https://doi.org/10.1126/sciadv.abl3840} {\bibfield  {journal} {\bibinfo
  {journal} {Sci. Adv.}\ }\textbf {\bibinfo {volume} {7}},\ \bibinfo {pages}
  {eabl3840} (\bibinfo {year} {2021})}\BibitemShut {NoStop}%
\bibitem [{\citenamefont {Soshenko}\ \emph {et~al.}(2021)\citenamefont
  {Soshenko}, \citenamefont {Bolshedvorskii}, \citenamefont {Rubinas},
  \citenamefont {Sorokin}, \citenamefont {Smolyaninov}, \citenamefont
  {Vorobyov},\ and\ \citenamefont {Akimov}}]{soshenko_nuclear_2021}%
  \BibitemOpen
  \bibfield  {author} {\bibinfo {author} {\bibfnamefont {V.~V.}\ \bibnamefont
  {Soshenko}}, \bibinfo {author} {\bibfnamefont {S.~V.}\ \bibnamefont
  {Bolshedvorskii}}, \bibinfo {author} {\bibfnamefont {O.}~\bibnamefont
  {Rubinas}}, \bibinfo {author} {\bibfnamefont {V.~N.}\ \bibnamefont
  {Sorokin}}, \bibinfo {author} {\bibfnamefont {A.~N.}\ \bibnamefont
  {Smolyaninov}}, \bibinfo {author} {\bibfnamefont {V.~V.}\ \bibnamefont
  {Vorobyov}},\ and\ \bibinfo {author} {\bibfnamefont {A.~V.}\ \bibnamefont
  {Akimov}},\ }\bibfield  {title} {\bibinfo {title} {Nuclear {{Spin Gyroscope}}
  based on the {{Nitrogen Vacancy Center}} in {{Diamond}}},\ }\href
  {https://doi.org/10.1103/PhysRevLett.126.197702} {\bibfield  {journal}
  {\bibinfo  {journal} {Phys. Rev. Lett.}\ }\textbf {\bibinfo {volume} {126}},\
  \bibinfo {pages} {6} (\bibinfo {year} {2021})}\BibitemShut {NoStop}%
\bibitem [{\citenamefont {Wang}\ \emph
  {et~al.}(2023{\natexlab{a}})\citenamefont {Wang}, \citenamefont {Barr},
  \citenamefont {Tang}, \citenamefont {Chen}, \citenamefont {Li}, \citenamefont
  {Xu}, \citenamefont {Stasiuk}, \citenamefont {Li},\ and\ \citenamefont
  {Cappellaro}}]{wang_characterizing_2023}%
  \BibitemOpen
  \bibfield  {author} {\bibinfo {author} {\bibfnamefont {G.}~\bibnamefont
  {Wang}}, \bibinfo {author} {\bibfnamefont {A.~R.}\ \bibnamefont {Barr}},
  \bibinfo {author} {\bibfnamefont {H.}~\bibnamefont {Tang}}, \bibinfo {author}
  {\bibfnamefont {M.}~\bibnamefont {Chen}}, \bibinfo {author} {\bibfnamefont
  {C.}~\bibnamefont {Li}}, \bibinfo {author} {\bibfnamefont {H.}~\bibnamefont
  {Xu}}, \bibinfo {author} {\bibfnamefont {A.}~\bibnamefont {Stasiuk}},
  \bibinfo {author} {\bibfnamefont {J.}~\bibnamefont {Li}},\ and\ \bibinfo
  {author} {\bibfnamefont {P.}~\bibnamefont {Cappellaro}},\ }\bibfield  {title}
  {\bibinfo {title} {Characterizing temperature and strain variations with
  qubit ensembles for their robust coherence protection},\ }\href
  {https://doi.org/10.1103/PhysRevLett.131.043602} {\bibfield  {journal}
  {\bibinfo  {journal} {Phys. Rev. Lett.}\ }\textbf {\bibinfo {volume} {131}},\
  \bibinfo {pages} {043602} (\bibinfo {year} {2023}{\natexlab{a}})}\BibitemShut
  {NoStop}%
\bibitem [{\citenamefont {Zhao}\ \emph {et~al.}(2022)\citenamefont {Zhao},
  \citenamefont {Shen}, \citenamefont {Ji},\ and\ \citenamefont
  {Huang}}]{zhao_inertial_2022}%
  \BibitemOpen
  \bibfield  {author} {\bibinfo {author} {\bibfnamefont {L.}~\bibnamefont
  {Zhao}}, \bibinfo {author} {\bibfnamefont {X.}~\bibnamefont {Shen}}, \bibinfo
  {author} {\bibfnamefont {L.}~\bibnamefont {Ji}},\ and\ \bibinfo {author}
  {\bibfnamefont {P.}~\bibnamefont {Huang}},\ }\bibfield  {title} {\bibinfo
  {title} {Inertial measurement with solid-state spins of nitrogen-vacancy
  center in diamond},\ }\href {https://doi.org/10.1080/23746149.2021.2004921}
  {\bibfield  {journal} {\bibinfo  {journal} {Adv. Phys.: X}\ }\textbf
  {\bibinfo {volume} {7}},\ \bibinfo {pages} {2004921} (\bibinfo {year}
  {2022})}\BibitemShut {NoStop}%
\bibitem [{\citenamefont {Wang}\ \emph {et~al.}(2024)\citenamefont {Wang},
  \citenamefont {Nguyen},\ and\ \citenamefont
  {Cappellaro}}]{wang_hyperfine-enhanced_2023}%
  \BibitemOpen
  \bibfield  {author} {\bibinfo {author} {\bibfnamefont {G.}~\bibnamefont
  {Wang}}, \bibinfo {author} {\bibfnamefont {M.-T.}\ \bibnamefont {Nguyen}},\
  and\ \bibinfo {author} {\bibfnamefont {P.}~\bibnamefont {Cappellaro}},\
  }\href@noop {} {\bibinfo {title} {Hyperfine-enhanced gyroscope based on
  solid-state spins}} (\bibinfo {year} {2024}),\ \Eprint
  {https://arxiv.org/abs/2401.01334} {arXiv:2401.01334 [quant-ph]} \BibitemShut
  {NoStop}%
\bibitem [{\citenamefont {Jaskula}\ \emph {et~al.}(2019)\citenamefont
  {Jaskula}, \citenamefont {Saha}, \citenamefont {Ajoy}, \citenamefont
  {Twitchen}, \citenamefont {Markham},\ and\ \citenamefont
  {Cappellaro}}]{jaskula_cross-sensor_2019}%
  \BibitemOpen
  \bibfield  {author} {\bibinfo {author} {\bibfnamefont {J.-C.}\ \bibnamefont
  {Jaskula}}, \bibinfo {author} {\bibfnamefont {K.}~\bibnamefont {Saha}},
  \bibinfo {author} {\bibfnamefont {A.}~\bibnamefont {Ajoy}}, \bibinfo {author}
  {\bibfnamefont {D.}~\bibnamefont {Twitchen}}, \bibinfo {author}
  {\bibfnamefont {M.}~\bibnamefont {Markham}},\ and\ \bibinfo {author}
  {\bibfnamefont {P.}~\bibnamefont {Cappellaro}},\ }\bibfield  {title}
  {\bibinfo {title} {Cross-{{Sensor Feedback Stabilization}} of an {{Emulated
  Quantum Spin Gyroscope}}},\ }\href
  {https://doi.org/10.1103/PhysRevApplied.11.054010} {\bibfield  {journal}
  {\bibinfo  {journal} {Phy. Rev. Applied}\ }\textbf {\bibinfo {volume} {11}},\
  \bibinfo {pages} {054010} (\bibinfo {year} {2019})}\BibitemShut {NoStop}%
\bibitem [{\citenamefont {Doherty}\ \emph {et~al.}(2013)\citenamefont
  {Doherty}, \citenamefont {Manson}, \citenamefont {Delaney}, \citenamefont
  {Jelezko}, \citenamefont {Wrachtrup},\ and\ \citenamefont
  {Hollenberg}}]{doherty_nitrogen-vacancy_2013}%
  \BibitemOpen
  \bibfield  {author} {\bibinfo {author} {\bibfnamefont {M.~W.}\ \bibnamefont
  {Doherty}}, \bibinfo {author} {\bibfnamefont {N.~B.}\ \bibnamefont {Manson}},
  \bibinfo {author} {\bibfnamefont {P.}~\bibnamefont {Delaney}}, \bibinfo
  {author} {\bibfnamefont {F.}~\bibnamefont {Jelezko}}, \bibinfo {author}
  {\bibfnamefont {J.}~\bibnamefont {Wrachtrup}},\ and\ \bibinfo {author}
  {\bibfnamefont {L.~C.}\ \bibnamefont {Hollenberg}},\ }\bibfield  {title}
  {\bibinfo {title} {The nitrogen-vacancy colour centre in diamond},\ }\href
  {https://doi.org/10.1016/j.physrep.2013.02.001} {\bibfield  {journal}
  {\bibinfo  {journal} {Phys. Rep.}\ }\textbf {\bibinfo {volume} {528}},\
  \bibinfo {pages} {1} (\bibinfo {year} {2013})}\BibitemShut {NoStop}%
\bibitem [{\citenamefont {Chen}\ \emph {et~al.}(2015)\citenamefont {Chen},
  \citenamefont {Hirose},\ and\ \citenamefont
  {Cappellaro}}]{chen_measurement_2015}%
  \BibitemOpen
  \bibfield  {author} {\bibinfo {author} {\bibfnamefont {M.}~\bibnamefont
  {Chen}}, \bibinfo {author} {\bibfnamefont {M.}~\bibnamefont {Hirose}},\ and\
  \bibinfo {author} {\bibfnamefont {P.}~\bibnamefont {Cappellaro}},\ }\bibfield
   {title} {\bibinfo {title} {Measurement of transverse hyperfine interaction
  by forbidden transitions},\ }\href
  {https://doi.org/10.1103/PhysRevB.92.020101} {\bibfield  {journal} {\bibinfo
  {journal} {Phys. Rev. B}\ }\textbf {\bibinfo {volume} {92}},\ \bibinfo
  {pages} {020101} (\bibinfo {year} {2015})}\BibitemShut {NoStop}%
\bibitem [{\citenamefont {Sangtawesin}\ \emph {et~al.}(2016)\citenamefont
  {Sangtawesin}, \citenamefont {McLellan}, \citenamefont {Myers}, \citenamefont
  {Jayich}, \citenamefont {Awschalom},\ and\ \citenamefont
  {Petta}}]{sangtawesin_hyperfine-enhanced_2016}%
  \BibitemOpen
  \bibfield  {author} {\bibinfo {author} {\bibfnamefont {S.}~\bibnamefont
  {Sangtawesin}}, \bibinfo {author} {\bibfnamefont {C.~A.}\ \bibnamefont
  {McLellan}}, \bibinfo {author} {\bibfnamefont {B.~A.}\ \bibnamefont {Myers}},
  \bibinfo {author} {\bibfnamefont {A.~C.~B.}\ \bibnamefont {Jayich}}, \bibinfo
  {author} {\bibfnamefont {D.~D.}\ \bibnamefont {Awschalom}},\ and\ \bibinfo
  {author} {\bibfnamefont {J.~R.}\ \bibnamefont {Petta}},\ }\bibfield  {title}
  {\bibinfo {title} {Hyperfine-enhanced gyromagnetic ratio of a nuclear spin in
  diamond},\ }\href {https://doi.org/10.1088/1367-2630/18/8/083016} {\bibfield
  {journal} {\bibinfo  {journal} {New J. Phys.}\ }\textbf {\bibinfo {volume}
  {18}},\ \bibinfo {pages} {083016} (\bibinfo {year} {2016})}\BibitemShut
  {NoStop}%
\bibitem [{\citenamefont {Oon}\ \emph {et~al.}(2022)\citenamefont {Oon},
  \citenamefont {Tang}, \citenamefont {Hart}, \citenamefont {Olsson},
  \citenamefont {Turner}, \citenamefont {Schloss},\ and\ \citenamefont
  {Walsworth}}]{oon_ramsey_2022-1}%
  \BibitemOpen
  \bibfield  {author} {\bibinfo {author} {\bibfnamefont {J.~T.}\ \bibnamefont
  {Oon}}, \bibinfo {author} {\bibfnamefont {J.}~\bibnamefont {Tang}}, \bibinfo
  {author} {\bibfnamefont {C.~A.}\ \bibnamefont {Hart}}, \bibinfo {author}
  {\bibfnamefont {K.~S.}\ \bibnamefont {Olsson}}, \bibinfo {author}
  {\bibfnamefont {M.~J.}\ \bibnamefont {Turner}}, \bibinfo {author}
  {\bibfnamefont {J.~M.}\ \bibnamefont {Schloss}},\ and\ \bibinfo {author}
  {\bibfnamefont {R.~L.}\ \bibnamefont {Walsworth}},\ }\bibfield  {title}
  {\bibinfo {title} {Ramsey envelope modulation in {{NV}} diamond
  magnetometry},\ }\href {https://doi.org/10.1103/PhysRevB.106.054110}
  {\bibfield  {journal} {\bibinfo  {journal} {Phys. Rev. B}\ }\textbf {\bibinfo
  {volume} {106}},\ \bibinfo {pages} {054110} (\bibinfo {year}
  {2022})}\BibitemShut {NoStop}%
\bibitem [{SOM()}]{SOM}%
  \BibitemOpen
  \href@noop {} {}\bibinfo {howpublished} {See Supplemental Material for
  details.}\BibitemShut {Stop}%
\bibitem [{\citenamefont {Chen}\ \emph {et~al.}(2018)\citenamefont {Chen},
  \citenamefont {Sun}, \citenamefont {Saha}, \citenamefont {Jaskula},\ and\
  \citenamefont {Cappellaro}}]{chen_protecting_2018}%
  \BibitemOpen
  \bibfield  {author} {\bibinfo {author} {\bibfnamefont {M.}~\bibnamefont
  {Chen}}, \bibinfo {author} {\bibfnamefont {W.~K.~C.}\ \bibnamefont {Sun}},
  \bibinfo {author} {\bibfnamefont {K.}~\bibnamefont {Saha}}, \bibinfo {author}
  {\bibfnamefont {J.-C.}\ \bibnamefont {Jaskula}},\ and\ \bibinfo {author}
  {\bibfnamefont {P.}~\bibnamefont {Cappellaro}},\ }\bibfield  {title}
  {\bibinfo {title} {Protecting solid-state spins from a strongly coupled
  environment},\ }\href {https://doi.org/10.1088/1367-2630/aac542} {\bibfield
  {journal} {\bibinfo  {journal} {New J. Phys.}\ }\textbf {\bibinfo {volume}
  {20}},\ \bibinfo {pages} {063011} (\bibinfo {year} {2018})}\BibitemShut
  {NoStop}%
\bibitem [{\citenamefont {Tang}\ \emph {et~al.}(2023)\citenamefont {Tang},
  \citenamefont {Barr}, \citenamefont {Wang}, \citenamefont {Cappellaro},\ and\
  \citenamefont {Li}}]{tang_first-principles_2023}%
  \BibitemOpen
  \bibfield  {author} {\bibinfo {author} {\bibfnamefont {H.}~\bibnamefont
  {Tang}}, \bibinfo {author} {\bibfnamefont {A.~R.}\ \bibnamefont {Barr}},
  \bibinfo {author} {\bibfnamefont {G.}~\bibnamefont {Wang}}, \bibinfo {author}
  {\bibfnamefont {P.}~\bibnamefont {Cappellaro}},\ and\ \bibinfo {author}
  {\bibfnamefont {J.}~\bibnamefont {Li}},\ }\bibfield  {title} {\bibinfo
  {title} {First-{{Principles Calculation}} of the {{Temperature-Dependent
  Transition Energies}} in {{Spin Defects}}},\ }\href
  {https://doi.org/10.1021/acs.jpclett.3c00314} {\bibfield  {journal} {\bibinfo
   {journal} {J. Phys. Chem. Lett.}\ }\textbf {\bibinfo {volume} {14}},\
  \bibinfo {pages} {3266} (\bibinfo {year} {2023})}\BibitemShut {NoStop}%
\bibitem [{\citenamefont {Lourette}\ \emph {et~al.}(2022)\citenamefont
  {Lourette}, \citenamefont {Jarmola}, \citenamefont {Acosta}, \citenamefont
  {Birdwell}, \citenamefont {Budker}, \citenamefont {Doherty}, \citenamefont
  {Ivanov},\ and\ \citenamefont {Malinovsky}}]{lourette_temperature_2022}%
  \BibitemOpen
  \bibfield  {author} {\bibinfo {author} {\bibfnamefont {S.}~\bibnamefont
  {Lourette}}, \bibinfo {author} {\bibfnamefont {A.}~\bibnamefont {Jarmola}},
  \bibinfo {author} {\bibfnamefont {V.~M.}\ \bibnamefont {Acosta}}, \bibinfo
  {author} {\bibfnamefont {A.~G.}\ \bibnamefont {Birdwell}}, \bibinfo {author}
  {\bibfnamefont {D.}~\bibnamefont {Budker}}, \bibinfo {author} {\bibfnamefont
  {M.~W.}\ \bibnamefont {Doherty}}, \bibinfo {author} {\bibfnamefont
  {T.}~\bibnamefont {Ivanov}},\ and\ \bibinfo {author} {\bibfnamefont {V.~S.}\
  \bibnamefont {Malinovsky}},\ }\href@noop {} {\bibinfo {title} {Temperature
  {{Sensitivity}} of \$\^\{14\}\textbackslash
  mathrm\{\vphantom\}{{NV}}\vphantom\{\}\$ and \$\^\{15\}\textbackslash
  mathrm\{\vphantom\}{{NV}}\vphantom\{\}\$ {{Ground State Manifolds}}}}
  (\bibinfo {year} {2022}),\ \Eprint {https://arxiv.org/abs/2212.12169}
  {2212.12169 [cond-mat, physics:physics, physics:quant-ph]} \BibitemShut
  {NoStop}%
\bibitem [{\citenamefont {Liu}\ \emph {et~al.}(2019)\citenamefont {Liu},
  \citenamefont {Ajoy},\ and\ \citenamefont {Cappellaro}}]{liu_nanoscale_2019}%
  \BibitemOpen
  \bibfield  {author} {\bibinfo {author} {\bibfnamefont {Y.-X.}\ \bibnamefont
  {Liu}}, \bibinfo {author} {\bibfnamefont {A.}~\bibnamefont {Ajoy}},\ and\
  \bibinfo {author} {\bibfnamefont {P.}~\bibnamefont {Cappellaro}},\ }\bibfield
   {title} {\bibinfo {title} {Nanoscale {{Vector}} dc {{Magnetometry}} via
  {{Ancilla}}-{{Assisted Frequency Up}}-{{Conversion}}},\ }\href
  {https://doi.org/10.1103/PhysRevLett.122.100501} {\bibfield  {journal}
  {\bibinfo  {journal} {Phys. Rev. Lett.}\ }\textbf {\bibinfo {volume} {122}},\
  \bibinfo {pages} {100501} (\bibinfo {year} {2019})}\BibitemShut {NoStop}%
\bibitem [{\citenamefont {Wang}\ \emph
  {et~al.}(2023{\natexlab{b}})\citenamefont {Wang}, \citenamefont {Li},
  \citenamefont {Tang}, \citenamefont {Li}, \citenamefont {Madonini},
  \citenamefont {Alsallom}, \citenamefont {Calvin~Sun}, \citenamefont {Peng},
  \citenamefont {Villa}, \citenamefont {Li},\ and\ \citenamefont
  {Cappellaro}}]{wang_manipulating_2023}%
  \BibitemOpen
  \bibfield  {author} {\bibinfo {author} {\bibfnamefont {G.}~\bibnamefont
  {Wang}}, \bibinfo {author} {\bibfnamefont {C.}~\bibnamefont {Li}}, \bibinfo
  {author} {\bibfnamefont {H.}~\bibnamefont {Tang}}, \bibinfo {author}
  {\bibfnamefont {B.}~\bibnamefont {Li}}, \bibinfo {author} {\bibfnamefont
  {F.}~\bibnamefont {Madonini}}, \bibinfo {author} {\bibfnamefont {F.~F.}\
  \bibnamefont {Alsallom}}, \bibinfo {author} {\bibfnamefont {W.~K.}\
  \bibnamefont {Calvin~Sun}}, \bibinfo {author} {\bibfnamefont
  {P.}~\bibnamefont {Peng}}, \bibinfo {author} {\bibfnamefont {F.}~\bibnamefont
  {Villa}}, \bibinfo {author} {\bibfnamefont {J.}~\bibnamefont {Li}},\ and\
  \bibinfo {author} {\bibfnamefont {P.}~\bibnamefont {Cappellaro}},\ }\bibfield
   {title} {\bibinfo {title} {Manipulating solid-state spin concentration
  through charge transport},\ }\href {https://doi.org/10.1073/pnas.2305621120}
  {\bibfield  {journal} {\bibinfo  {journal} {Proceedings of the National
  Academy of Sciences}\ }\textbf {\bibinfo {volume} {120}},\ \bibinfo {pages}
  {e2305621120} (\bibinfo {year} {2023}{\natexlab{b}})}\BibitemShut {NoStop}%
\end{thebibliography}%

\newpage
\clearpage
\setcounter{section}{0}
\setcounter{equation}{0}
\setcounter{figure}{0}
\setcounter{table}{0}
\setcounter{page}{1}
\makeatletter
\renewcommand{\theequation}{S\arabic{equation}}
\renewcommand{\thesection}{S\arabic{section}}
\renewcommand{\thefigure}{S\arabic{figure}}


\title{Supplemental Materials}
\maketitle
\onecolumngrid
\section{Hamiltonian}
\subsection{Derivation of the hyperfine enhancement factor}

The $^{15}$NV center is modeled by an electronic spin ($S = 1$) coupled to the native $^{15}$N nuclear spin ($I = 1/2$). The electron-nuclear spin-mixing augments the gyromagnetic ratio of the nuclear spin, leading to an enhancement factor $\alpha_{m_s} \equiv \gamma_{N, eff}/\gamma_{N, bare}$ dependent on the electronic spin state and applied magnetic field,  that increases dramatically around GSLAC \cite{chen_measurement_2015, sangtawesin_hyperfine-enhanced_2016}.  Here, we carry out the derivation for the precise expressions for the enhancement factors in each NV spin manifold.

The ground state Hamiltonian of the NV center is given by $H = \Vec{S}\cdot \bold{D}\cdot \Vec{S} + \gamma_e \Vec{B} \cdot \Vec{S} + \gamma_n \Vec{B} \cdot \Vec{I} + \Vec{S} \cdot \bold{A} \cdot \Vec{I}$.  Under an applied magnetic field $B_z$, the Hamiltonian can be written as a sum of secular $H_{||}$ and non-secular terms $H_{\perp}$:
\begin{equation}
    H_{||} = DS_z^2 + \gamma_e B_z S_z + \gamma_n B_z I_z + A_{zz}S_zI_z
\end{equation}
\begin{equation}
    H_{\perp} = A_{\perp}S_xI_x + A_{\perp}S_yI_y = \frac{A_{\perp}}{2}(S_+I_- + S_-I_+).
\end{equation}
The raising and lowering operators are defined as $S_{\pm} = S_x \pm i S_y$ and $I_{\pm} = I_x \pm i I_y$, where $S$ and $I$ are the electron spin-1 and nuclear spin-1/2 operators, respectively. The transverse coupling $A_{\perp}$ couples the $H_{||}$ eigenstates $|m_S, m_I \rangle = |+1, -\frac{1}{2}\rangle \leftrightarrow |0, +\frac{1}{2}\rangle$ and  $|0, -\frac{1}{2}\rangle \leftrightarrow |-1, +\frac{1}{2}\rangle$,  which define the zero-quantum (ZQ) subspaces.  The total Hamiltonian can be diagonalized by rotating the ZQ subspaces with the unitary transformation $U_{ZQ} = e^{-i(\theta^-\sigma_y^- + \theta^+\sigma_y^+)}$ where the rotation operators are given by $\sigma_y^+ = i(|+1, -\frac{1}{2}\rangle \langle 0, +\frac{1}{2}| - |0, +\frac{1}{2}\rangle \langle +1, -\frac{1}{2}|)$ and $\sigma_y^- = i(|0, -\frac{1}{2}\rangle \langle -1, +\frac{1}{2}| - |-1, +\frac{1}{2}\rangle \langle 0, -\frac{1}{2}|)$ and the rotation angles are given by 
\begin{equation}
    \tan(2 \theta^+) = \frac{2A_{\perp}}{D + \gamma_eB_z - \gamma_NB_z-A_{zz}/2}
\end{equation}
\begin{equation}
    \tan(2 \theta^-) = \frac{-2A_{\perp}}{D - \gamma_eB_z + \gamma_NB_z - A_{zz}/2}
\end{equation}

An additional transverse magnetic field in the NV frame gives an interaction Hamiltonian $H_x = B_x (\gamma_e S_x + \gamma_N I_x)$ (for simplicity here and in the main  paper, we consider only $B_x$, but a magnetic field component $B_y$ would see an equivalent enhancement).  Under the ZQ unitary transformation, the field is rotated to $\hat{H}_x = U_{ZQ}H_xU_{ZQ}^{\dagger}$.  Keeping only the nuclear spin terms, the effective Hamiltonian for the nuclear spin reads:
\begin{equation}
    \hat{H}_{x}^I = \gamma_N(\alpha_{+1}|+1\rangle \langle +1| + \alpha_{0}|0\rangle \langle 0| + \alpha_{-1}|-1\rangle \langle -1|),
\end{equation}
which defines the gyromagnetic ratio enhancement factors for the three electron spin states:  
\begin{align}
    \alpha_{+1} &= \cos(\theta^+) +  \frac{\gamma_e}{\gamma_n}\sin(\theta^+) \\
    \alpha_{0} &= \cos(\theta^+)\cos(\theta^-) - \frac{\gamma_e}{\gamma_n}\sin(\theta^+ - \theta^-) \\
    \alpha_{-1} &= \cos(\theta^-) - \frac{\gamma_e}{\gamma_n}\sin(\theta^-). 
\end{align}

\begin{figure}[h]
\includegraphics[width=0.5\textwidth]{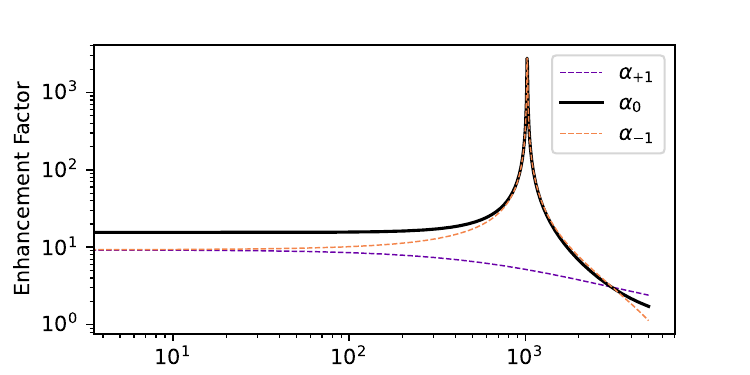}
\caption{\label{fig_supp:enhancement}
Predicted nuclear gyromagnetic enhancement factors in each NV electronic spin manifold $m_S$ as a function of $B_x$.  The derivation theoretically predicts a maximum finite enhancement ratio near GSLAC with $\alpha \approx \frac{\gamma_e}{\sqrt{2} \gamma_N}$. For the hyperfine-enhanced gyroscope, we consider the $m_S = 0$ manifold in particular, whose enhancement factor is shown in black.}
\end{figure}

Thus, for an applied magnetic field $\Vec{B} = (B_x, 0, B_z)$, we obtain the effective nuclear spin Hamiltonian for each electronic spin manifold $m_S$ presented in Eq. (\ref{eq:H_I}). The enhancement factor as a function of magnetic field is presented in Fig.~\ref{fig_supp:enhancement}, which in particular shows the extreme enhancement around the ground-state level anti-crossing.

\section{Experimental details}
\subsection{Setup}
The homoepitaxial nitrogen-doped layer was grown using plasma-enhanced chemical vapor deposition (PE-CVD). A 70 $\mu$m functional layer was grown on an $\langle 100\rangle$-oriented single-crystal diamond seed. After growth, the sample was electron irradiated at 1 MeV with a total dose of 3.1$\times10^{18}$ e$^-$/cm$^2$. Finally, the sample was annealed in a tube furnace at a temperature of 800~$\degree$C for several hours. 

The experimental setup used in this work is the same one in our previous work~\cite{wang_manipulating_2023}. By focusing a 0.5~W 532~nm laser to a broad area of about 200~$\mu$m diameter, more than $10^{10}$ spins are addressed simultaneously. The microwave and radio frequency control is applied through a printed circuit board with a 1~mm diameter loop. The fluorescence is collected by a photodiode through an objective with a numerical aperture of 0.5. The same setup also has a side-collection option to measure fluorescence emitted from the sides of the diamond, enabling compact gyroscope design for commercial applications in the future. A pair of permanent magnets (K\&J Magnetics DY08-N52) with three-dimensional translation and rotation degrees of freedom are used to apply a magnetic field of about 239~G along the NV axis (when perfectly aligned).

\subsection{Nuclear spin polarization and readout}
\begin{figure*}
\includegraphics[width=0.8\textwidth]{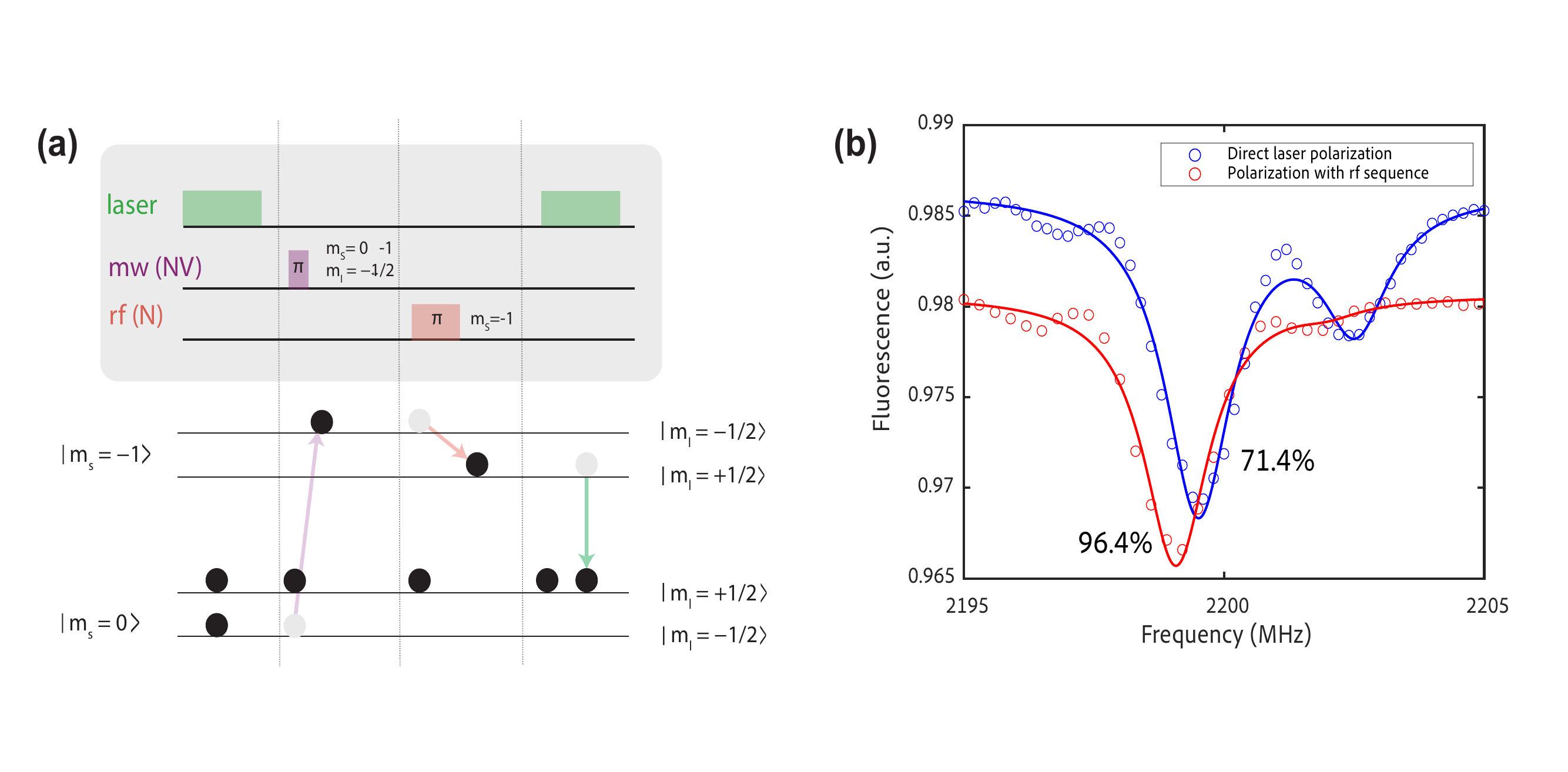}
\caption{\label{fig:pol_sequence} \textbf{Nuclear polarization} (a) Nuclear spin polarization sequence (b) Room-temperature ODMR signal of the NV for $m_s = 0 \leftrightarrow -1$ transition comparing nuclear spin polarization using ESLAC and polarization sequence.  Direct laser polarization through ESLAC yields $\approx 71 \%$ polarization, while the rf polarization sequence polarizes the nuclear spin to more than $96\%$. }
\end{figure*}
The NV electron spin state is optically pumped to the $m_s = 0$ sublevel of the ground state 
and the nuclear spin is polarized by transferring polarization from the NV electronic spin to the nuclear spins.  There are several methods to accomplish this: we can set the static magnetic field close to the electronic spin level crossing for either the ground or excited state of the NV center such that the electron-nuclear spin flip-flop can transfer the polarization from electronic spin to the nuclear spin, or we can use a polarization sequence combining selective microwave/rf and laser pulses for population transfer and electronic spin state re-initialization. A detailed discussion is as follows. 

The nuclear spin can be optically polarized with the electron spin.  When a magnetic field $B \approx 500G$ is applied along the axis of the NV center, the $m_s = 0$ and $m_s = -1$ sublevels of the excited state become nearly degenerate, and near this excited-state level anticrossing (ESLAC), strong hyperfine coupling in the excited state allows energy-conserving electron-nuclear spin flip-flop processes to occur between the coupled electron-nuclear spin states: $|m_S = 0, m_I = -\frac{1}{2}\rangle \leftrightarrow |m_S = -1, m_I = +\frac{1}{2}\rangle$.  Under ESLAC, the system is preferentially polarized to $| m_S = 0, m_I = +\frac{1}{2} \rangle$ because this state is not affected by mixing in the excited states and thus does not pass through the dark singlet states. We align our magnetic field $B = 239G$ along the NV axis with less than $0.3 \degree$ misalignment.  Using high laser power, we are able to polarize the nuclear spin to $|m_I = +\frac{1}{2}\rangle$ with a polarization of about $70\%$ (Fig. \ref{fig:pol_sequence}(b)).

The direct laser polarization method requires the magnetic field to be close to ESLAC or GSLAC and the field alignment is perfect. To achieve better polarization of the nuclear spin at a field lower than 500G without perfect alignment, we can also polarize the nuclear spin to  $|m_I = +\frac{1}{2}\rangle$ using a polarization sequence that transfers the electronic spin polarization, shown in Fig. \ref{fig:pol_sequence}(a).  First, we transfer the population from the $|m_S = 0, m_I=-\frac12\rangle$ spin state to $|m_S = -1, m_I=-\frac12\rangle$ using a selective microwave $\pi$ pulse with a Rabi frequency of 1~MHz that is smaller than the 3~MHz hyperfine splitting.  Then, we apply a selective rf $\pi$ pulse that transfers the $|m_S = -1, m_I = -\frac{1}{2}\rangle$ population to $|m_S = -1, m_I = +\frac{1}{2}\rangle$.  Finally, a green laser pulse is applied to transfer the population to the $m_S = 0$ electronic spin level while preserving the nuclear spin state.  This allows us to polarize the system to $|m_S = 0, m_I = +\frac{1}{2} \rangle$. This procedure is robust against magnetic field strength and misalignment, allowing us to reach near unity nuclear spin polarization $\approx 96 \%$, which we use to perform the experiments in the paper. 

Optical readout of the nuclear spin state after gyroscope rotation sensing is accomplished directly with MW mapping pulses shown in Fig. $\ref{fig:2}$.

\subsection{Characterization of Coherence Times}
\begin{figure*}
\includegraphics[width=0.99\textwidth]{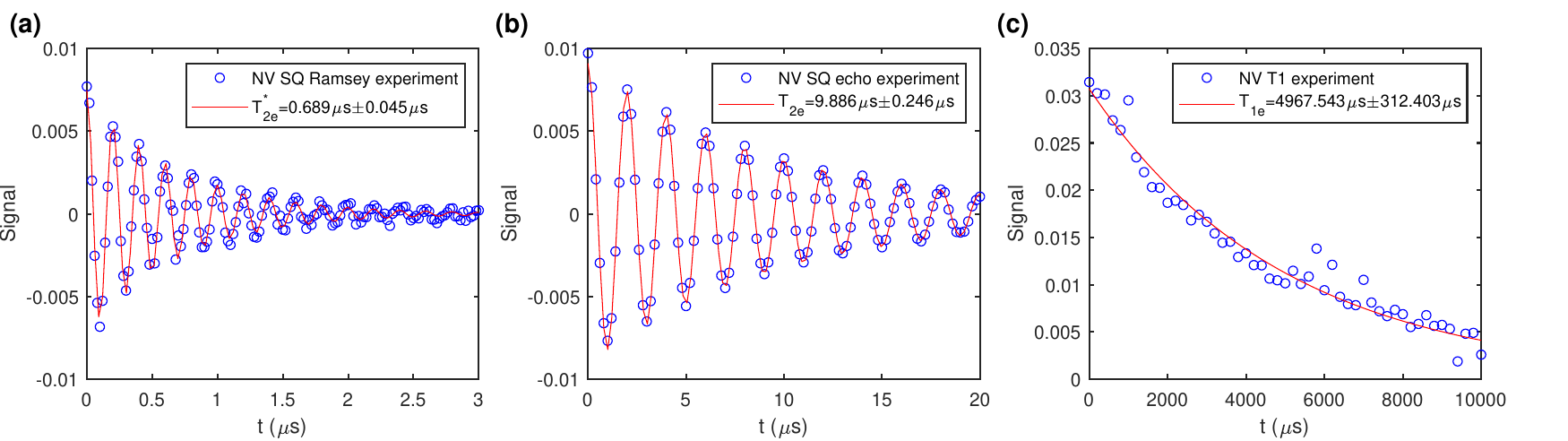}
\caption{\label{fig:NVT1T2T2} \textbf{NV electronic spin coherence $T_{2e}^*,T_{2e}$ and lifetime $T_{1e}$ measurements.} }
\end{figure*}

The NV electronic spin Ramsey dephasing time $T_{2e}^*$, spin echo coherence time $T_{2e}$ and spin relaxation time $T_{1e}$ are characterized in Fig.~\ref{fig:NVT1T2T2}.

\subsection{Misalignment raw data}
\begin{figure*}
\includegraphics[width=0.8\textwidth]{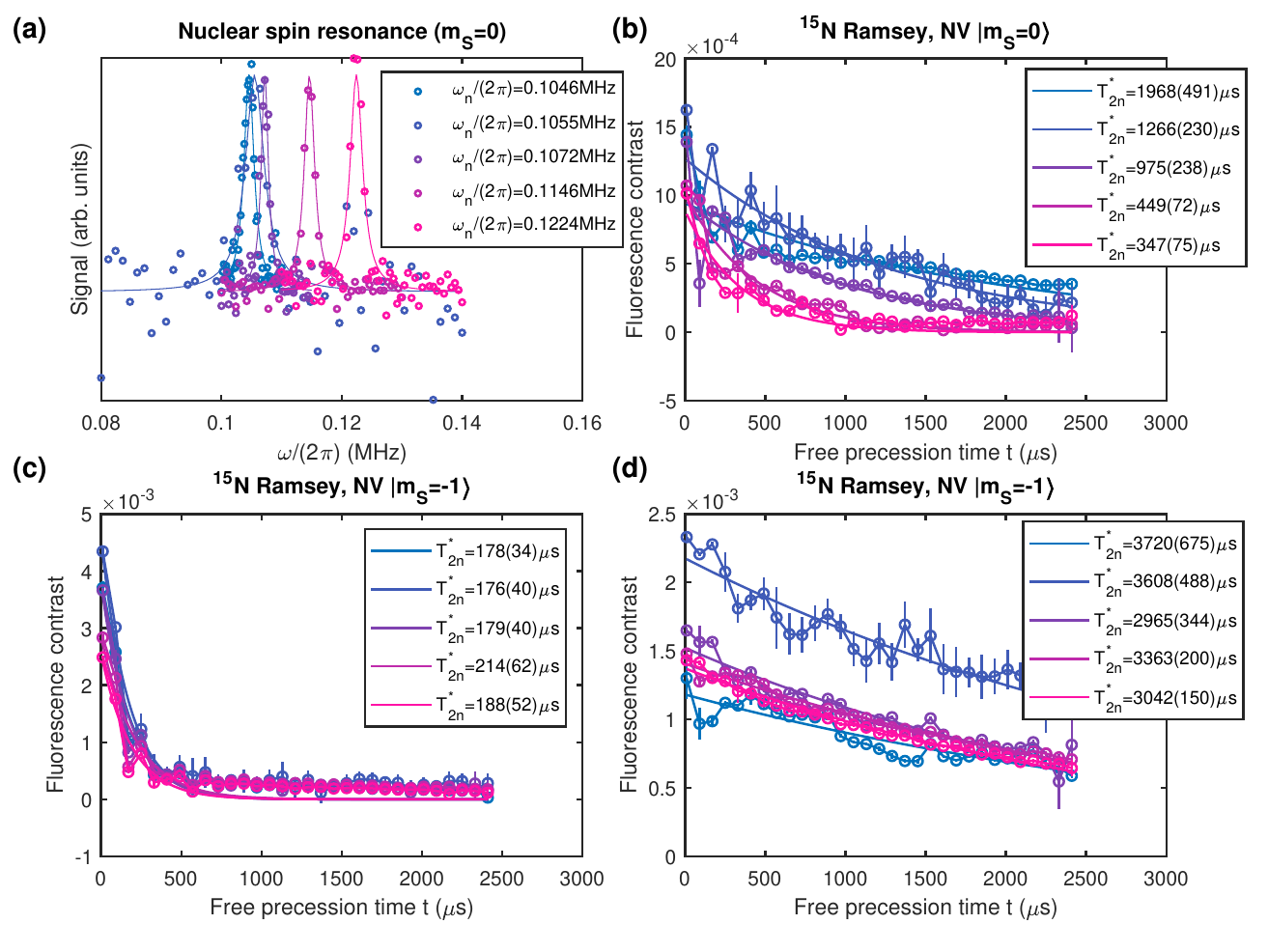}
\caption{\label{fig:misalignRaw} \textbf{Misalignment data.} (a) Measurement of the nuclear spin transition frequencies in the electronic spin $\ket{m_S=0}$ under different misalignment conditions. (b) Measurement of  nuclear spin dephasing in the electronic spin $\ket{m_S=0}$ manifold. Different data numbers correspond to the different misalignment conditions calibrated in (a), the same rules apply to (c) and (d). (c) Measurement of nuclear spin dephasing in the electronic spin $\ket{m_S=-1}$ manifold. (d) Measurement of nuclear spin dephasing under the protection of electronic spin DQ $\pi$ pulse. }
\end{figure*}

The raw data of the nuclear spin dephasing time under different misalignment conditions are shown in Fig.~\ref{fig:misalignRaw}. To calibrate the misalignment, we measure the nuclear spin transition frequencies $\omega_{measure}$ in the electronic spin $\ket{m_S=0}$ manifold (Fig.~\ref{fig:misalignRaw}(a)). By comparing the measured transition frequencies with the theoretically predicted ones $\omega_{measure} =\omega(\theta)$, we can  obtain the misalignment angles $\theta=\omega^{-1}(\omega_{measure})$. The function $\omega(\theta)$ is calculated using the exact diagonalization of the system Hamiltonian under a magnetic field $B=239$~G misaligned from the NV $z$ axis by an angle $\theta$. 

The nuclear spin dephasing times are measured by sweeping the phase of the last $\pi/2$ nuclear spin control pulse at each evolution time $t$ and the fitting of the oscillation amplitude in these experiments are shown in Figs.~\ref{fig:misalignRaw}(b-d). The oscillation amplitudes are then fit to exponential decay function $S(t)=ce^{-t/T_{2n}^*}$ to extract the dephasing times $T_{2n}^*$ shown in the legends. In these dephasing experiments, the electronic spin is first prepared to $\ket{m_S=-1}$ before applying nuclear spin $\pi/2$ pulse to prepare the superposition state $(\ket{+1/2}+\ket{-1/2})/\sqrt{2}$. For the measurement in Fig.~\ref{fig:misalignRaw}(b), the electronic spin state is transferred to $\ket{m_S=0}$ with a non-selective $\pi$ pulse (with a frequency applied at the middle of the two resonances corresponding to two nuclear spin hyperfine states and the $\pi$ pulse duration is 0.1~$\mu$s, corresponding to a Rabi frequency of 5~MHz larger than the hyperfine splitting) after preparing the superposition state.  For the measurement in Fig.~\ref{fig:misalignRaw}(d), a double-quantum (DQ) $\pi$ pulse is applied in the middle of the evolution, which is composed of three non-selective $\pi$ pulses of transition $\ket{m_S=0}\leftrightarrow\ket{m_S=-1}$, $\ket{m_S=0}\leftrightarrow\ket{m_S=+1}$, $\ket{m_S=0}\leftrightarrow\ket{m_S=-1}$.
For both (b) and (d) cases, the same SQ and DQ $\pi$ pulse is applied again before the last $\pi/2$ pulse on the nuclear spin and selective $\pi$ pulse for population readout.

\subsection{Temperature shift}
\begin{figure*}
\includegraphics[width=0.8\textwidth]{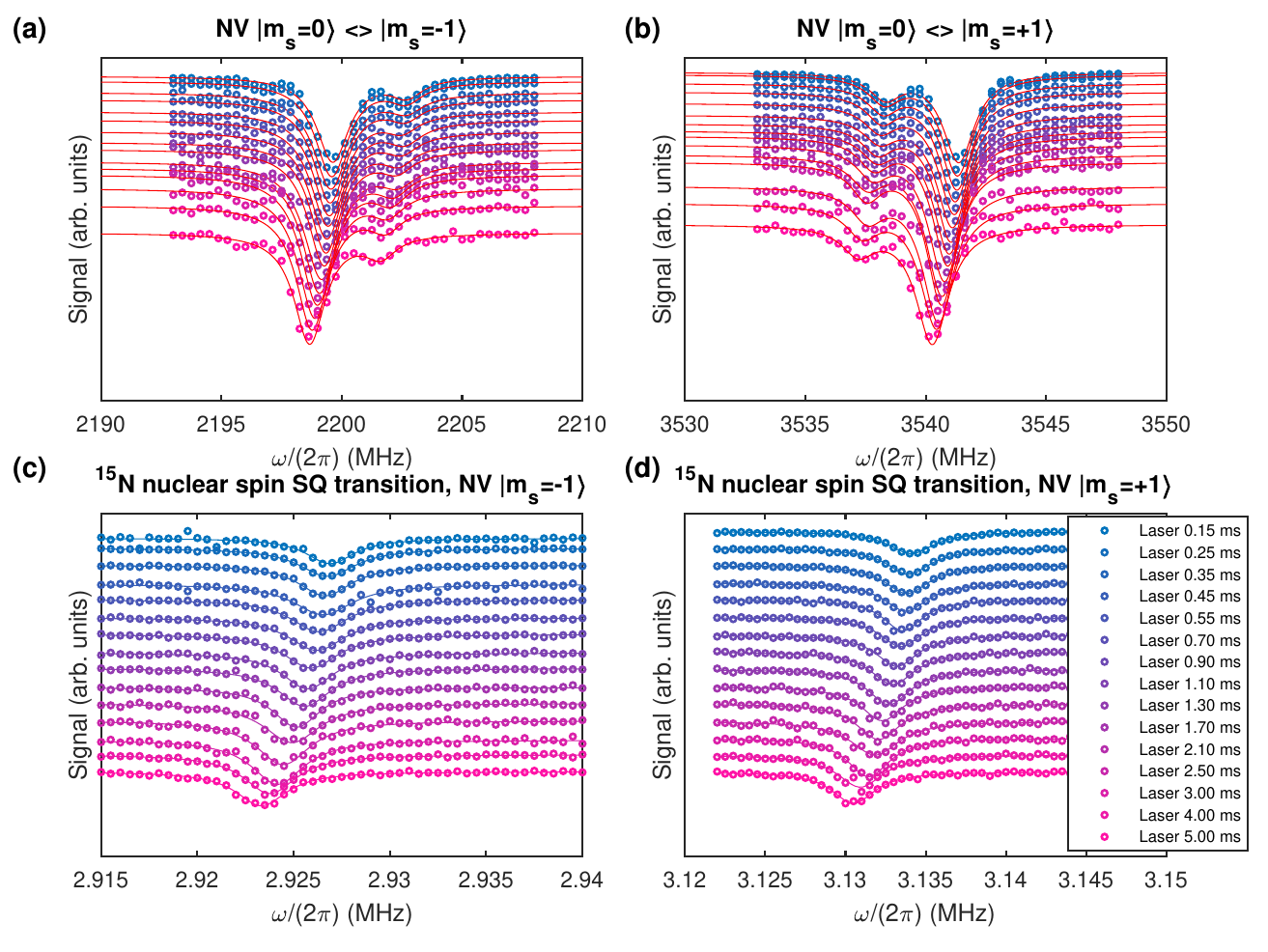}
\caption{\label{fig:TShiftsRaw} \textbf{Temperature shift raw data.} (a,b) NV electron spin resonance measurement under different laser duration. (c,d) Nuclear spin magnetic resonance frequency measurement in the $\ket{m_S=\mp1}$ electronic spin manifold. }
\end{figure*}

\begin{figure*}
\includegraphics[width=0.99\textwidth]{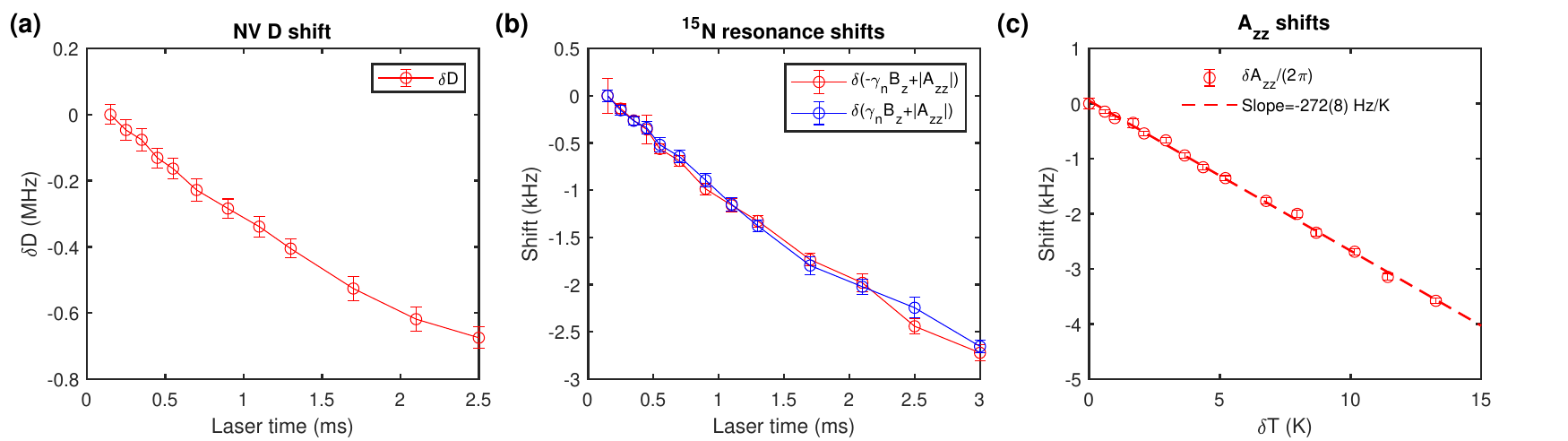}
\caption{\label{fig:TShifts} \textbf{Temperature shift data.} (a) Shifts of NV zero-field splitting $D$ as a function of temperature. (c) Shifts of the nuclear spin transition frequencies in the $\ket{m_S=\pm1}$ electronic spin manifolds. (d) Shifts of the longitudinal hyperfine interaction as a function of the temperature change. }
\end{figure*}

The temperature shift of hyperfine interaction is characterized by varying the laser illumination duration and measuring the nuclear spin transition frequencies in both of the two electronic spin state manifolds $|m_S=\pm1\rangle$. The measured results are shown in Fig.~\ref{fig:TShiftsRaw}. The NV electronic spin resonance frequencies are measured to calibrate the temperature shift using zero-field splitting (the average of the two SQ transition frequencies) under different laser illumination times (Fig.~\ref{fig:TShifts}(a)). 
The average of the two nuclear spin transition frequencies (electronic spin state manifolds $|m_S=\pm1\rangle$) under the same illumination conditions (Fig.~\ref{fig:TShifts}(b)) is used to calculate the shift in the hyperfine interaction as shown in Fig.~\ref{fig:TShifts}(c).

\end{document}